# Studies on Fabrication of Ag/Tl$_2$Ba$_2$Ca$_2$Cu$_3$O$_{10}$/CdSe Heterostructures with Electrochemical Technique

**P M  Shirage, D D Shivagan and S H Pawar***

School of Energy Studies, Department of Physics Shivaji University,
Kolhapur-416 004, INDIA.

E-mail pawar_s_h@yahoo.com, shpawar_phy@unishivaji.ac.in

**Abstract**

The Ag/Tl$_2$Ba$_2$Ca$_2$Cu$_3$O$_{10}$/CdSe heterostructure was successfully fabricated by complex bath electrodeposition technique for the first time. The nucleation growth mechanism by studying the chronoamperographs during the deposition process both for Tl$_2$Ba$_2$Ca$_2$Cu$_3$O$_{10}$ (Tl-2223) and CdSe layers. The formation of the heterostructure with non-diffusive interfaces was confirmed by X-ray diffraction and Scanning Electron Microscopy (SEM). The crystalline size determined by using FWHM and Scheerer's formula, were found 33 nm and 25 nm respectively for Tl-2223 ans CdSe films.

The Tl$_2$Ba$_2$Ca$_2$Cu$_3$O$_{10}$ electrodeposited on Ag-substrate has shown the superconducting transition temperature $T_c$ at 116.5 K and $J_c$ = 2.1 x 10$^3$ A/cm$^2$ and was found to improve after the CdSe over layer deposition The improvement in $T_c$ and $J_c$ was also found after the Red He-Ne laser irradiation onto the film. The improvement in $T_c$ and $J_c$ of Tl2Ba2Ca2Cu3O10 in heterostructure and with laser irradiation are discussed at length in the paper.

**\* Author for correspondence**

# 1 Introduction

One of the most expected technological directions for the high temperature superconductors (HTSCs) is the hetero-epitaxial growth of the multilayer structures. This may be able to integrate a variety of the other technologies with superconductivity at liquid nitrogen (Liq. $N_2$ 77 K ) and will help for the realization of the most of cryogenic devices based on the superconducting thin films. This development of the multilayer techniques will be crucial. One of such multilayer structures is a hybrid device which include non-hysteric Josephson junctions, in which the coupling between the HTSC layer is provided by a semiconductor layer.

Since the electronic state in the semiconductor can be easily controlled by electrostatic potential, these superconductor-semiconductor-superconductor or superconductor-semiconductor junction with a gate configuration are very useful for the systematic study of the quantum transport. These hybrid multi-layered heterostructures also posses a possible application in optoelectronics, cryoelectronics, optical detectors, electron coolers and in three terminal devices. Also the fabrication of layered structures of the superconducting and normal-metallic or insulating films present interesting challenges for a wide range of the electronic applications. Similarly the metal-superconductor-semiconductor heterostructure has created a new platform in view of physicist research and its application.

Some progress has been made in the growth of multi-layers with HTSC thin films on Si [1] and GaAs [2,3] as a substrate for the HTSC films by molecular beam epitaxy (MBE) technique. A major problem in using the Si and GaAs as the substrate for the HTSC films is the substrate-film inter-diffusion and the formation of micro-cracks in the films due to relative difference in thermal coefficient of expansion between the semiconducting and HTSC films [4] during the fabrication of the heterostructure by MBE. The heterostructures between Bi-2212 (BSCCO) and GaAs formed by depositing GaAs onto BSCCO by molecular beam expitaxy technique showed great enhancement in $T_c$ from 71 K to 83 K [4]. Rao *et al.* [5] have grown InAs on Tl-2212 (TlBaCaCuO) superconducting film. Pawar *et al.* [6,7] have also observed anomalous behavior in I-V characteristics of Ag/BSCCO/CdSe heterostructure at 250 K, suggesting a possibility of increasing $T_c$ as high as 250 K.

The superconducting properties can be enhanced by two ways. Firstly the superconducting properties such as $T_c$ and $J_c$ values can be improved by illumination of superconducting film with light or by photodoping. This photodoping could lead, not only to an increase in conductivity, but could also make possible to enhance the $T_c$ and $J_c$ [7,8]. Secondly, the photosensitive semiconducting layer can be deposited on the superconductor and after illumination such a structure with light, the electron-hole pairs are generated at the interface. The electrons get pumped in semiconductor while the holes be pushed into the superconducting layers [ ]. Thus, increase in carrier concentration in HTSC leads to increase in $T_c$.

Hence there is a great need of to fabricate variety of the metal Superconductor and Semiconductor heterostructures. In present investigation, the attempts have been made to fabricate the Ag/$Tl_2Ba_2Ca_2Cu_3O_{10}$/CdSe abrupt heterostructure by electrodeposition technique and reported first time.

## 2 Experimental Procedure

The room temperature electrochemical synthesis of $Tl_2Ba_2Ca_2Cu_3O_{10}$[ ], Hg-Ba-Ca-CuO[ ], Y-Ba-CuO[ ]and $MgB_2$ [ ] this films are reported elsewhere.

The heterostructure of the superconducting Tl-2223 and semiconducting CdSe was fabricated onto silver substrate by using electrodeposition technique. The composition and optimization were parameters of the Tl-2223 superconducting films on Ag-substrate are elsewhere [ Shirage P M, Ph.D. thesis Shivaji University, Kolhapur December 2002 and IJC ] .

The bath composition for the synthesis of CdSe on Ag-substrate and Ag/$_{Tl2Ba2Ca2Cu3O10}$ films were 10 mM CdSO$_4$ and 50 mM SeO$_2$ prepared in double distilled water. The pH of the bath was 3.3, i.e. acidic in nature. The Princeton Applied Research Perkin Elmer VersaStat-II model 250/270 electrochemical software interfaced with the P-III PC and the three electrode cell were employed for the electrochemical experiment. For the deposition of CdSe, Ag-substrate and Tl-2223 films were used as the working electrodes, saturated calomel electrode (SCE) as the reference electrode and graphite as the counter electrode.

SEM images were obtained with a SEM probed with EDAX on Philips Model XL-30. The structural characterization were characterized by X-ray diffraction (XRD) using Phillips PW-3710 diffractometer using CuK$_\alpha$ radiations. The particle size was calculated using full width at half maximum (FWHM) and by applying Scherrer's formula. Four probe type contacts with equal area were made onto deposit to measure the electrical properties. The red He-Ne laser ( $\lambda$ = 632.8 nm, power = 2 mW and photon energy = 1.95 eV ) was used to irradiate the heterostructures.

**3 Results and Discussion:**

The probable problem arising in heterostructures fabrication are lattice mismatch, thermal coefficient of expansion and inter-diffusion of the substrate into films. Some of the semiconductors with their lattice parameters and their lattice mismatch with Tl$_2$Ba$_2$Ca$_2$Cu$_3$O$_{10}$ are listed in following Table 1.

**Table 1** *List of the semiconductors with the % lattice mismatch*

*with* $Tl_2Ba_2Ca_2Cu_3O_{10}$ *superconductor*

| Semiconductor | Lattice parameter *'a'* (Å) | % Lattice Mismatch with $Tl_2Ba_2Ca_2Cu_3O_{10}$ ( a = 3.85 Å ) |
|---|---|---|
| InAs | 6.06 | 44.60 |
| GaAs | 5.65 | 37.89 |
| CdSe | 4.20 | 8.69 |
| CdTe | 4.57 | 17.10 |
| ZnTe | 6.101 | 31.84 |
| Si | 5.4307 | 34.06 |
| Ge | 5.65 | 37.89 |

From Table 1, it is seen that CdSe has lower *% lattice mismatch* compare to other and it has good photosensitive property. Because of the CdSe semiconductors suitable band gap (1.74 eV) and high photosensitivity in the visible range of the solar spectrum, these materials can be used for the low cost applications [19]. So it was decided to use semiconductor CdSe in heterostructure fabrication.

The present investigation deals with the fabrication of Ag/Tl-2223/CdSe and Ag/CdSe/Tl-2223 heterostructures. These geometries were formed by two ways.

In the first case, superconducting Tl-2223 system onto Ag-substrate was synthesized as discussed earlier [  ] (Electrochemically oxidized for 28 min, optimum oxidation period). Then subsequently CdSe layer was deposited onto the superconducting films, to form a geometry as Ag/Tl-2223/CdSe heterostructure.

In the second case, CdSe was deposited onto silver substrate and then subsequently followed by Tl-Ba-Ca-Cu alloy deposition and electrochemical oxidation for 28 min.

Following are some steps, which were adopted for the formation of these geometries by electrochemical technique.

### 3.1 Linear Sweep Voltammetry (LSV)

### 3.1.1 LSV For Cd and Se Deposition onto Ag-Substrate

The linear sweep voltammetry for the deposition of Cd and Se from the aqueous solution of $CdSO_4$ and $SeO_2$ onto Ag substrate were carried out for the estimation of the deposition potentials of Cd and Se separately, shown in Fig. 1(a) and (b) respectively.

From Fig. 1(a) it is observed that the reduction potential of Cd is at $-750$ mV $vs$ SCE. The reaction taking place at this potential is,

$$Cd^{2+} + 2e^- \rightarrow Cd_{(s)} \tag{1}$$

At this potential, gray-blackish deposition was observed onto the Ag-substrate.

Fig. 1(b) shows linear sweep voltammetry curve for deposition of the Se onto the Ag-substrate. The curve shows the gradual increase in current with voltage. So it is not possible to locate the exact deposition potential. This was optimized by taking the deposition at various potentials. The optimized deposition potential was the $-500$ mV $vs$ SCE. The reaction taking place at potential $-500$ mV $vs$ SCE is,

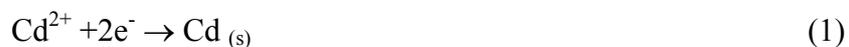

$$SeO_2 + H_2O \rightarrow H_2SeO^-_{3\ (aq)} \tag{2}$$

$$H_2SeO^-_{3(aq)} + 5H^+ + 4e^- \rightarrow Se_{(s)} + 3H_2O \tag{3}$$

At the deposition potential of − 500 mV *vs* SCE the brown-yellowish deposit was observed on the Ag-substrate.

From the LSV studies the deposition potential of the Cd and Se were found at − 750 mV *vs* SCE and  - 500 mV *vs* SCE respectively.

### 3.1.2  Linear Sweep Voltammetry for Deposition of Cdse onto Ag-Substrate and Ag/Tl-2223 Film

In Fig. 2, curves (a) and (b) shows the LSV for CdSe deposition onto Ag-substrate and Ag/Tl-2223 films. The curve Fig. 2  (a) shows a peak at the  − 540 mV  *vs*  SCE where as the Fig. 2 (b) shows the peak at − 700 mV *vs* SCE. The peaks at these are may be due to the initiation of the deposition process of CdSe. These potentials gives the possible deposition potentials of CdSe. Some under and over potentials are required to optimize the deposition potential of CdSe. The optimized potential for CdSe deposition onto Ag-substrate and Ag/Tl-2223 films was  - 600 mV *vs* SCE.

The over all mechanism taking place during the deposition of the CdSe can be summarized as [20],

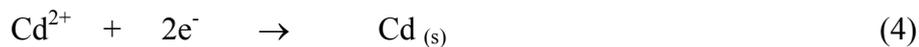
$$Cd^{2+} + 2e^- \rightarrow Cd_{(s)} \tag{4}$$

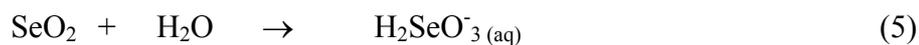
$$SeO_2 + H_2O \rightarrow H_2SeO^-_{3\,(aq)} \tag{5}$$

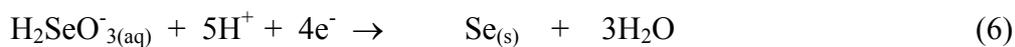
$$H_2SeO^-_{3(aq)} + 5H^+ + 4e^- \rightarrow Se_{(s)} + 3H_2O \tag{6}$$

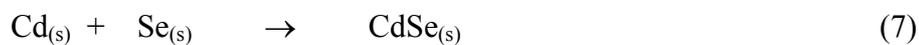
$$Cd_{(s)} + Se_{(s)} \rightarrow CdSe_{(s)} \tag{7}$$

At potential – 600 mV *vs* SCE the smoke-blackish deposition was observed onto Ag-substrate and Ag/Tl-2223 films.

It is observed that the reduction potential of the CdSe on Ag-substrate and Ag/Tl-2223 films is in between the deposition potential of the Se and Cd deposition on Ag-substrate when estimated separately onto Ag-substrate. The reduction potential of the CdSe in between the Cd and Se was also reported earlier Yesugade *et al.* [21] and Lade *et al.* [19].

### 3.2.Chromoampermetry

This technique can be utilized to study the kinetics and nucleation process during the electrodeposition. The typical chronoamperographs recorded for the deposition of CdSe onto Ag-substrate and Ag/Tl-2223 films at potential (-600 mV) are shown in Fig. 3, curves (a) and (b) respectively. From Fig. 3 curve (a) and (b) it has been observed that the current density for CdSe deposition onto Ag-substrate and Ag/Tl-2223 films decrease drastically within the first few seconds and is attributed to the formation of a double layer at the electrode-electrolyte interface, which causes an increase in resistance. For CdSe deposition onto Ag-substrate, from Fig.3 curve (a), beyond this drastic decrease, the current density starts to increase within a few minutes and then remains constant afterwards. This may be due to Coulombic repulsion, some ions are repelled towards the cathode and the discharge of double layer may takes place giving a rise in current. After this small increase, current density remains constant afterwards due to steady flow of ions across the electrode.

The current density of CdSe deposition onto Ag/Tl-2223 film was smaller compared to current density of CdSe deposition onto Ag-substrate. The decrease in current density during CdSe deposition on Ag/Tl-2223 film attributes to the presence of oxide layer, which is more resistive than the metallic Ag-substrate. Ag metal is more conducting than Ag/Tl-2223. The current density of CdSe deposition onto Ag and Ag/Tl-2223 system were found to be 6 mA/cm$^2$ and 3 mA/cm$^2$, respectively.

Determination of the nucleation process was achieved by analyzing the portion of the current–transient and then comparing the curve to dimension less theoretical curve for the instantaneous and progressive nucleation process,

$$(I/I_{max})^2 = 1.9542(t/t_{max})^{-1}\{1-exp[-1.2564(t/t_{max})]\}^2 \qquad (8)$$

$$(I/I_{max})^2 = 1.2254(t/t_{max})^{-1}\{1-exp[-2.3367(t/t_{max})]\}^2 \qquad (9)$$

where $I_{max}$ and $t_{max}$ are the maximum current and time in current-time transient. Fig. 5.4(a) and Fig. 4 (b) shows a selection of dimensionless experimental curve and their corresponding theoretical curves that results from equation (8) and (9). From Fig. 4 (a) and (b) it is seen that dimensionless plots agree favorably with a progressive nucleation process at potential - 600 mV *vs* SCE. Such progressive nucleation process is needed to occur the deposition of CdSe.

### 3.3 Thickness Measurements

The thickness of the thin film is important factor, which affects the electrical properties. The thickness was measured by weight-density difference method. The variation of CdSe thickness with the deposition period is shown in Fig. 5. Rate of deposition of CdSe onto Ag-substrate was 0.1 $\mu$m/min. onto Ag/Tl-2223 was 0.075 $\mu$m/min. The thickness of the CdSe deposition onto Ag-substrate and Ag/Tl-2223 film were found to be 2 $\mu$m and 1.5 $\mu$m respectively after 20 min. deposition. The rate of the CdSe deposition on Ag substrate is higher than the Ag/Tl-2223 film due to higher conductivity.

### 3.4 Deposition of Tl-Ba-Ca-Cu Alloy onto Ag/CdSe and Its Oxidation

The fabrication of $Tl_2Ba_2Ca_2Cu_3O_{10}$ superconductor on Ag/CdSe film also been tried. The bath composition for the deposition of the Tl-Ba-Ca-Cu alloy was the same as discussed in earlier [ ] and experimental procedure remains same, substrate being Ag/CdSe film.

The linear sweep voltammetry plot for the deposition of Tl-Ba-Ca-Cu alloy onto Ag/CdSe substrate is shown in Fig 6. Fig. 6 shows two peaks. The peak at $-1.4$ V $vs$ SCE, correspond to deposition of $Tl^+$, $Ba^{2+}$, $Ca^{2+}$ and $Cu^{2+}/Cu^{3+}$. The peak at $-0.5$V $vs$ SCE may correspond to under potential deposition of the $Tl^+$. Chronoamperometry plot at potential $-1.4$ V $vs$ SCE is shown in Fig. 7 (a). Chronoamperomtry show the fall in current after the maxima, which may be due to formation of the double layer at the electrode-electrolyte interface.

Fig. 7 (b) shows experimental plot of $(I/I_{max})^2$ $vs$ $(t/t_{max})$ along with the dimension less theoretical plots for instantaneous and progressive nucleation process. Fig. 7 (b) show the experimental plots are of the progressive nucleation process.

Deposition potential at potential $-1.4$ V $vs$ SCE gives the black deposit of the Tl-Ba-Ca-Cu alloy of onto Ag/CdSe. The optimized period of the Tl-Ba-Ca-Cu alloy deposition was 12 min. After 12 min., the deposit was found to peel off into the solution. The maximum thickness calculated from the weight-density difference method was found to be 2.3 $\mu$m with rate of deposition $\approx 0.19$ $\mu$m/min. The current density of the deposition of Tl-Ba-Ca-Cu

**alloy onto Ag/CdSe was found to be 1mA/cm$^2$, which is less than the current density for Tl-Ba-Ca-Cu alloy deposition on Ag-substrate.** ---- see whether the last is in continueation- ---- This may be due to the high resistance of CdSe than the Ag-substrate.

Then using the 1N KOH solution as the electrolytic bath, Ag/CdSe/Tl-Ba-Ca-Cu alloy as the working electrode and graphite as the counter electrode, the deposited films were oxidized. The LSV plot for the intercalation of the oxygen onto Ag/CdSe/Tl-Ba-Ca-Cu alloy film is as shown in Fig. 8. From this the oxidation potential was optimized to be + 700 mV *vs* SCE to intercalate oxygen into Tl-Ba-Ca-Cu alloy deposited onto Ag/CdSe deposition onto Ag/CdSe. The electrochemical oxidation was carried out for 28 min [  ]

### 3.5 X-Ray Diffraction (XRD) Studies

X-ray diffraction (XRD) was employed to characterize the structure and films phase composition. The XRD pattern of the CdSe is shown in Fig. 9 (a). The x-ray diffraction pattern of the CdSe was analyzed by using standard ASTM data [22] and indexed with hexagonal CdSe structure. The *a* and *c*-axis parameters were calculated by using the standard '*d*' spacing,

$$\frac{1}{d^2} = \frac{4(h^2 + hk + k^2)}{3 \ a^2} + \frac{l^2}{c^2} \qquad (10)$$

The presence of the (002) characteristics peak at $2\theta = 25.45^o$ along with (103) and (203) planes confirms the CdSe deposition onto Ag substrate. The lattice parameters (a and c) were calculated are 4.25 Å and 7.07 Å respectively. These parameters are closely agree with the standard reported data [23]. The crystalline size determined by full width at half maxima (FWHM) and Scheerer's formula is 25 nm. The presence of the characteristics (002) at

$2\theta = 25.40^{\circ}$ plane in the electrochemically deposited CdSe films on the stainless steel was also reported [22].

Fig. 9(b) shows the XRD pattern of Ag/Tl-2223/CdSe heterostructure. The XRD pattern was indexed Tl-2223 planes with the tetragonal lattice parameters and CdSe with the hexagonal lattice parameters.

The presence of major peaks (103), (00$\underline{14}$), (118), (00$\underline{18}$), (00$\underline{15}$), (11$\underline{14}$) and (10$\underline{23}$) confirms tetragonal superconductor Tl-2223 phase. The presence of major peak (002) along with (100), (110), (103), (201) and (203) confirms the hexagonal semiconductor CdSe. The particle sizes of the CdSe and Tl-2223 were calculated from FWHM and Scherrer formula and were found to be 25 nm and 33 nm respectively. The XRD pattern shows the presence of superconducting Tl-2223 and semiconducting CdSe. The calculated c-parameter of Tl-2223 is 35.57 Å and CdSe is 7.1 Å. These are in good agreement with the values reported earlier and confirm the formation of heterostructure. The formation of the nano-crystalline CdSe from XRD was also reported by Bawendi *et al.*[24] and Golan *et al.* [25].

The XRD pattern of the Ag /Tl-2223/CdSe heterostructure is shown Figure 9 (c). The XRD pattern of Ag /CdSe/$_{Tl2Ba2Ca2Cu3O10}$ heterostructure shows the (008), (00$\underline{14}$), (118), (200), (11$\underline{14}$) and (10$\underline{23}$) planes of Tl-2223 phases and (002), (103), (201) and (203) planes CdSe peaks with large number of unidentified impurity peaks. The unidentified impurity peaks were not indexed. The XRD pattern was indexed with the tetragonal lattice parameters for Tl-2223 phase and hexagonal lattice parameters for CdSe. The calculated c-parameter of Tl-2223 is 35.97 Å and CdSe is 7.4 Å. The calculated 'c' parameters are found more than the earlier report. This may is because of the heterostructure may not formed. The presence of impurity peaks indicated that the heterostructure is not formed. Hence this study was not carried out further.

*3.6 Scanning Electron Microscopy (SEM) Studies*

The SEM gives the key information about the surface and the most properties of materials are related to the surfaces. The top view of the $CdSe/Tl_2Ba_2Ca_2Cu_3O_{10}$ in $Ag/CdSe/Tl_2Ba_2Ca_2Cu_3O_{10}$ heterostructure is shown in Fig 10 (a). This shows two distinct regions clear dark and bright region, left side Tl-2223 superconductor and right side is CdSe semiconductor. The higher magnification images of Tl-2223 and CdSe are shown in Fig 10 (b) and 10 (c) respectively. When the images are higher magnified (10000x) it shows the highly dense grain image. The grain sizes of the individual materials were not determined by using the SEM as the grains are not clearly seen in image and crystallite size in the nm range. Both images show compact grains. No cracks and pinholes were observed.

*3.7 Electrical Properties*

The CdSe film of 1.5 μm thickness was deposited on 1.5 cm. length of the $Tl_2Ba_2Ca_2Cu_3O_{10}$ deposited films. $Tl_2Ba_2Ca_2Cu_3O_{10}$ films used here are single phase. The four equi-distance line contacts were made on the top of the heterostructures by using the air drying silver paint. The change in resistivity of film is measured as a function of temperature. These measurements were normalized to the respective value at 300 K and presented in Fig. 11 curve (a). The normal state resistivity of the heterostructure is $4.0 \times 10^3$ Ωcm. It is seen that resistivity remains constant for initial first fall of 25 K and then starts gradually increasing and remains flat at after 90 K. This represents that, although the resistivity of the superconductor decreases with temperature, the resistivity of semiconducting CdSe increases at low temperatures, and hence there is gradual increase in resistivity after 200 K. This may be due the fact that the rate of increase in resistance of the CdSe is more than the decrease in the resistance of the $Tl_2Ba_2Ca_2Cu_3O_{10}$. The resistivity continues to increase further and remains steady after 90 K. The fall in resistivity was expected at 116.5 K, the transition temperature of Tl-2223 used in this system. However, it is interesting to note that the resistivity of individual CdSe layer deposited onto Ag-substrate has resistivity of the order of $10^3$ $\Omega$-cm. ($4.0 \times 10^3$ $\Omega$ cm.) and that of $Tl_2Ba_2Ca_2Cu_3O_{10}$ deposited was 10 $\Omega$ cm.. Because of large difference in resistivity values and dominance of CdSe, the drop would not have appeared in resistivity.

This sample is irradiated with red He-Ne laser ( $\lambda$ = 6.328 nm, E = 1.95 eV and the power = 2 mW) for 2 hours irradiation. Fig. 11 curve (b) shows the resistivity measurements carried out during the laser irradiation. It is seen that the resistivity is decreased after the laser irradiation, but the nature of variation of resistivity with temperature remains similar to that of without irradiation state. The decrease in resistivity is attributed to increase in conductivity due to laser irradiation. The CdSe has the direct band gap 1.74 eV and in this experiment the sample was deliberately irradiated with the laser having the photon energy 1.95 eV. Due irradiation with laser which is having more photon energy that the band gap of CdSe, the electron-hole pair generates in the bulk of CdSe. Due to the junction between CdSe (n) and $Tl_2Ba_2Ca_2Cu_3O_{10}$(p), the electron hole pairs is do not allowed to recombine rather holes are transferred to the superconducting state where they can be further trapped into $CuO_2$ layer and hence there is increase in conductivity.

The steady state variation in the pattern at about 110 K might be due to the transformation of sandwiched $Tl_2Ba_2Ca_2Cu_3O_{10}$ into superconducting state. The contact resistivity values may be considered as the residual high contact resistant due to semiconducting CdSe.

Rao *et al.* [5] have developed InAs film with 3500 Å thickness about TlBaCaCuO(2212) film using molecular beam epitaxy techniques. They have performed the four point measurements of resistance (R) as a function of temperature (T) using press contacts and found that under lying TlBaCaCuO is superconducting at 100 K. Whereas the TlBaCaCuO film without InAs was superconducting at 106 K and $T_c$ is not significantly altered by the deposition of InAs. They attributed that the superconducting behaviour was achieved because of the low resistivity InAs as it was produced with the higher doping levels [5].

Masao Nakao [4] has observed the problem during measurement of resistivity of MgO/BSCCO/GaAs heterostructures. Hence, he has made slight change in geometry of the deposition and deposited GaAs on the middle of the BSCCO films. The contacts were made on the BSCCO and on the two sides of GaAs deposits. He observed the enhancement in superconducting transition from 71 K to 83 K by the deposition of GaAs on top of BSCCO. The increase in superconducting parameters were attributed to the fact that GaAs deposited on BSCCO prevents the loss of oxygen from BSCCO film during annealing [4].

In the present investigation, it was planed to implement the contact geometry as given by Masao Nakao[4] with some slight modifications in current electrodes so that it would be possible to collect all the carriers generated during the photo-irradiations.

CdSe was deposited onto the $Tl_2Ba_2Ca_2Cu_3O_{10}$ and contacts made are shown Fig. 12. The $Tl_2Ba_2Ca_2Cu_3O_{10}$ films were deposited on Ag-substrate with the dimensions of 1.5 cm x 3 cm The $Tl_2Ba_2Ca_2Cu_3O_{10}$ films were deposited with the dimensions of 1.5 cm x 1.5 cm onto silver substrate of dimensions 3 cm. x 1.5 cm. The four line contacts were made along the edges of the $Tl_2Ba_2Ca_2Cu_3O_{10}$ films as shown in Fig 5.12 living sufficient area of 0.8 cm x 1 cm for the deposition of CdSe. The sample area, except middle 0.8 cm x 1 cm was physically covered by the transparent tape and CdSe is then deposited for the optimize parameters.

Fig. 13 curve (a) shows the change in the resistance of Tl-Ba-Ca-CuO as a function of temperature without CdSe depositions. Sample shows zero resistivity at 116.5 K. The $J_c$ value measured at 77 K is $2.1 \times 10^3$ A/cm.$^2$ Fig. 13 curve (b) shows the resistivity of CdSe deposited onto the $Tl_2Ba_2Ca_2Cu_3O_{10}$. It shows the similar resistivity behavior but zero resistivity is achieved at 118.7 K, a slight improvement is resulted. The $J_c$ value measured is $2.3 \times 10^3$ A/cm.$^2$ at 77 K. The superconducting parameters here improved are small. Then increase in superconductors parameters is attributed to micro level dislocations induced by a lattice mismatch between the $Tl_2Ba_2Ca_2Cu_3O_{10}$ and CdSe at the interface [26]. It is further be attributed to the fact that some of the Se in CdSe is present in elemental form and is very sensitive to oxygen and forms $SeO_2$. When CdSe is deposited onto the $Tl_2Ba_2Ca_2Cu_3O_{10}$ oxide films the elemental Se forms the bound state with the oxygen present on the surface and physically the surface is said to be clean for superconductor. But due to bound states of Oxygen and CdSe due to formation $SeO_2$, the probability of formation of diffusive layer may takes place. This type of chemi-sorption of oxygen by elemental Se is discussed by Nair *et al*. [18].

This CdSe area of the samples is irradiated by red He-Ne laser with $\lambda$ = 632.8 nm, E = 1.95 eV and power p = 2 mW) for the 3 hours. The variations of the resistivity were measured as a function of time and are shown in Fig.13. It can be excitingly observed that the $T_c$ is further increased to 121 K and $J_c$ value measured at 77 K is $3.0 \times 10^3$ A/cm$^2$.

The $Tl_2Ba_2Ca_2Cu_3O_{10}$ sample used here showed 116.5 K transition temperature and the oxygen content is optimum. Hence, in this case the increasing superconducting parameters can only be attributed to the increase in carrier concentration when the sample was irradiated by laser having the energy greater than the band gap of semiconductor, it generates the carrier concentrations. The minimum energy gap of 1.74 eV of bulk CdSe suggest an onset of optical absorption at 714 nm. As prepared film shows the onset of optical absorption near 620 nm and hence an optical band gap of grater than 1.74 eV. Such increase in band gap of chemically deposited CdSe thin films, upto 0.5 eV higher than in single crystal samples has been reported before. This was explained in terms of a quantum size effect arising due to very small crystallite size of ~ 5 nm in the film. However, in our case the crystallite size measured by Scherrer formulae is 25 nm and hence the possibility of high gap is ruled out. However, we have irradiated the CdSe by red laser having photon energy = 1.95 eV, sufficiently greater than 1.74 eV. The presence of oxygen is notable in the surface layer, a common feature of metal chalcogenide thin films. The presence of adsorbed oxygen in the intergrain region improves the photosensitivity of the films.

Hence, the $SeO_2$ at the interface between Tl-2223 and CdSe acts as the trapping centers and the barrier potentials at interface allows the hole to transfer into Tl-2223 and could not get annihilated. Hence, in this way the hole concentration in the superconductors increases. This might be the reason of the increasing in superconductivity parameters, particularly the $J_c$ values.

Recently, *Foget et al* [27] have discussed the dislocation-induced superconductivity in novel superconducting-semiconducting superlattices (SL). The great body of the experimental data showed that the $T_c$ of the artificial multilayer as rule does not exceed the transition temperature of the superconducting films constituting SL [28]. As a single exception the semiconducting SL's PbTe/PbS and PbTe/SnTe reveling $T_c$ upto 6 K [29, 30] may be considered. The regular grids of the misfit dislocations on the interfaces of the epitaxial PbTe/PbS SL's is regarded as a phenomenon related to the superconductivity.

In the present investigation, the increase in superconducting parameters after deposition of CdSe and laser irradiation is attributed the micro level dislocations due to lattice mismatch, attachment of selenium to the oxygen present on the surface, creating a diffusion region in semiconductor side of the interface. Upon irradiation these $SeO_2$ particles acts as trapping centers and the photogenerated holes are transferred to the superconductor and could not be recombined due to the potential barrier of the interface. The holes transferred to the superconductors may be trapped into $CuO_2$ layers and due to increase in hole concentrations the superconducting parameters might have increased.

## 4   Conclusions

The Ag/$Tl_2Ba_2Ca_2Cu_3O_{10}$/CdSe heterostructure was successfully synthesized by electrodeposition technique. The planes of hexagonal CdSe and tetragonal Tl-2223 in the X-ray diffraction pattern confirms the formation of Ag/ $Tl_2Ba_2Ca_2Cu_3O_{10}$/CdSe heterostructure. The crystalline size determined by using FWHM and Scheerer's formulla were in nano-crystalline range. The $T_c$ and $J_c$ values were found to improve after the CdSe deposition onto Ag/ $Tl_2Ba_2Ca_2Cu_3O_{10}$. The improvement in $T_c$ and $J_c$ values also observed after laser irradiation and is attributed to increase in carrier concentration at the interface. The Ag/ $Tl_2Ba_2Ca_2Cu_3O_{10}$/CdSe heterostructure can be used to study the interfacial properties of the materials forming the junctions.

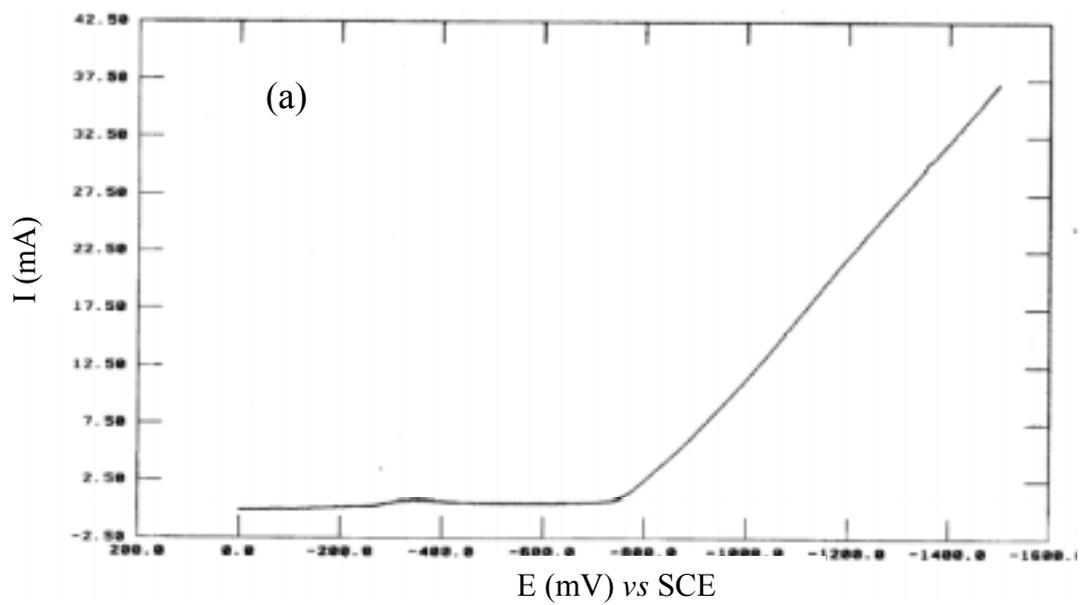

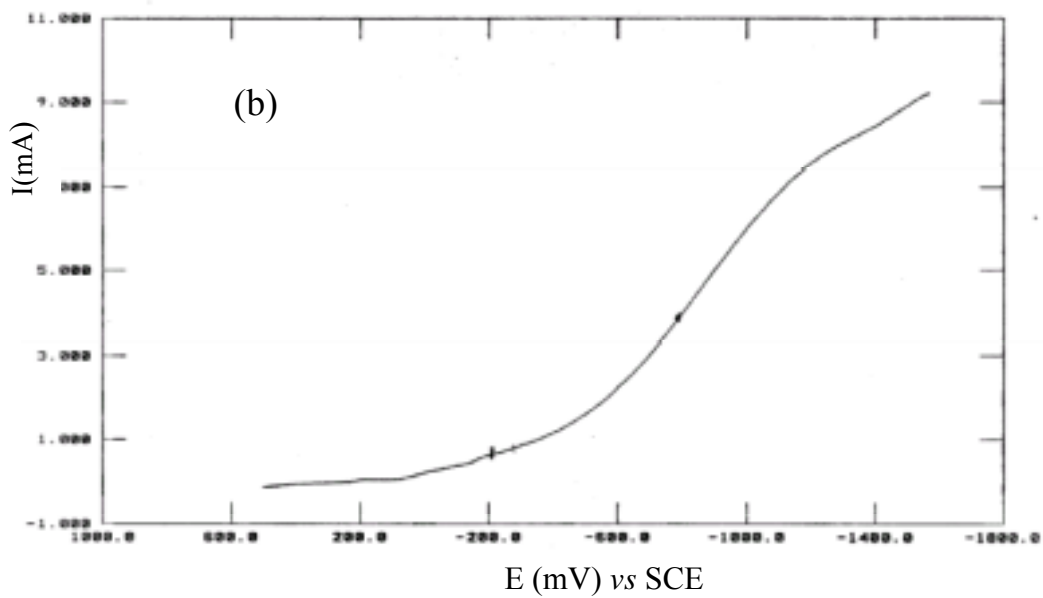

**Figure 1**    Linear sweep voltammograms for the deposition of
(a) Cd from the aqueous $CdSO_4$ bath and
(b) Se from aqueous $SeO_2$ bath

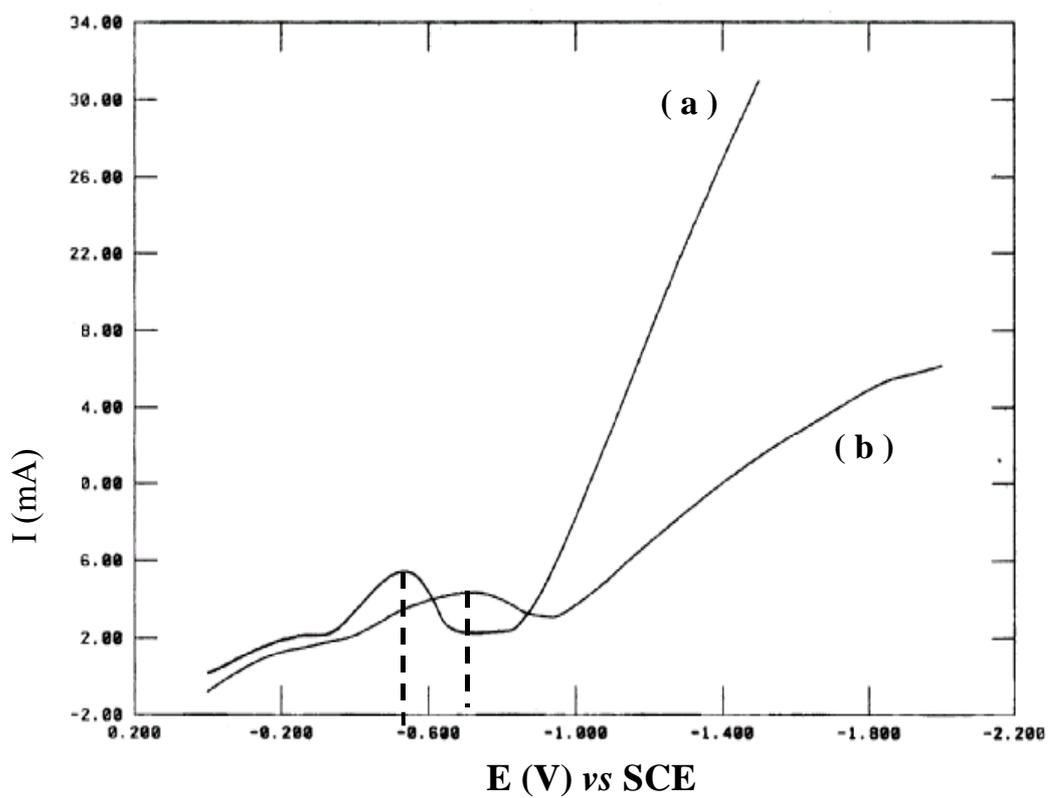

**Figure 2**    Linear sweep voltammograms for deposition of
CdSe on (a) Ag-substrate (b) Ag/Tl-2223 film

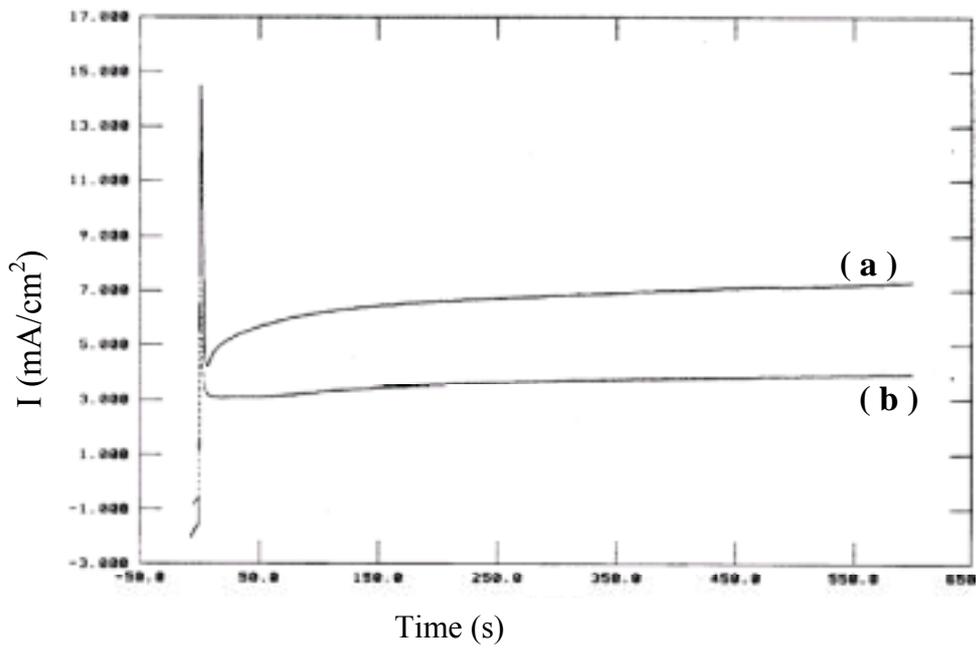

**Figure 3**    Chronoamperometry    plots    during    the deposition of CdSe on
(a) Ag-Substrate and  (b) Ag/Tl-2223 film

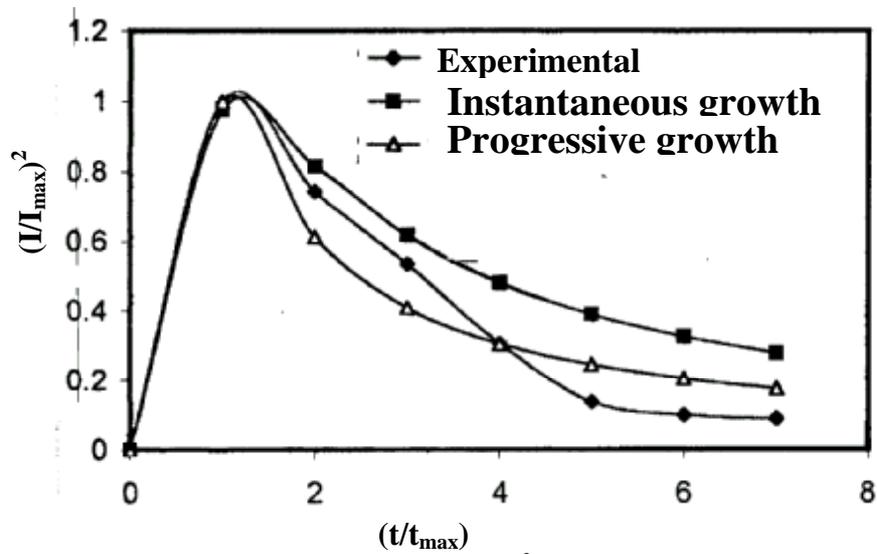

**Figure 4(a)** The plot of $(I/I_{max})^2$ vs $(t/t_{max})$ for the study of nucleation and growth mechanism during the deposition of CdSe onto Ag substrate at $-600$ mV *vs* SCE.

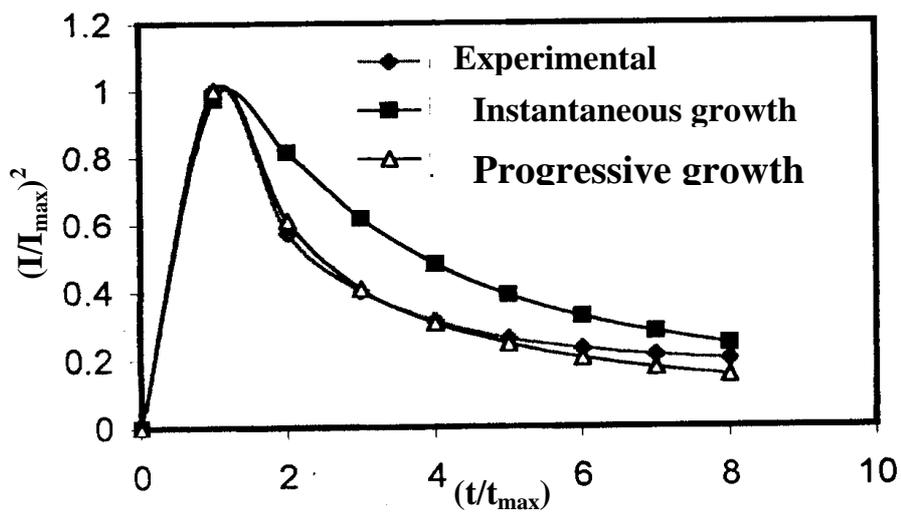

**Figure 4(b)** The plot of $(I/I_{max})^2$ vs $(t/t_{max})$ for the study of nucleation and growth mechanism during the deposition of CdSe onto Ag /Tl-Ba-Ca-CuO film at $-600$ mV *vs* SCE.

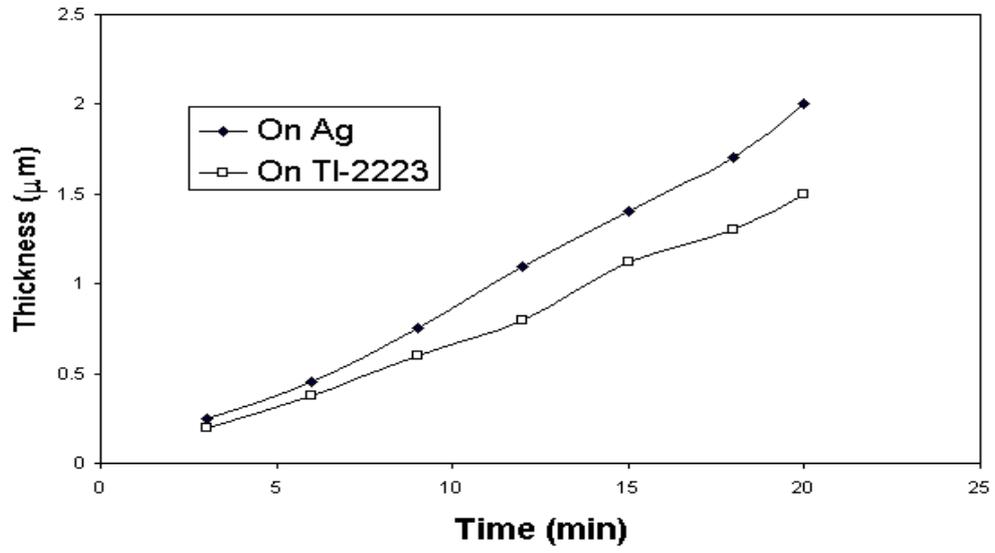

**Figure 5**     Variation of the CdSe thickness with time for the
deposition onto (a) Ag-substrate (b) Ag/Tl-2223

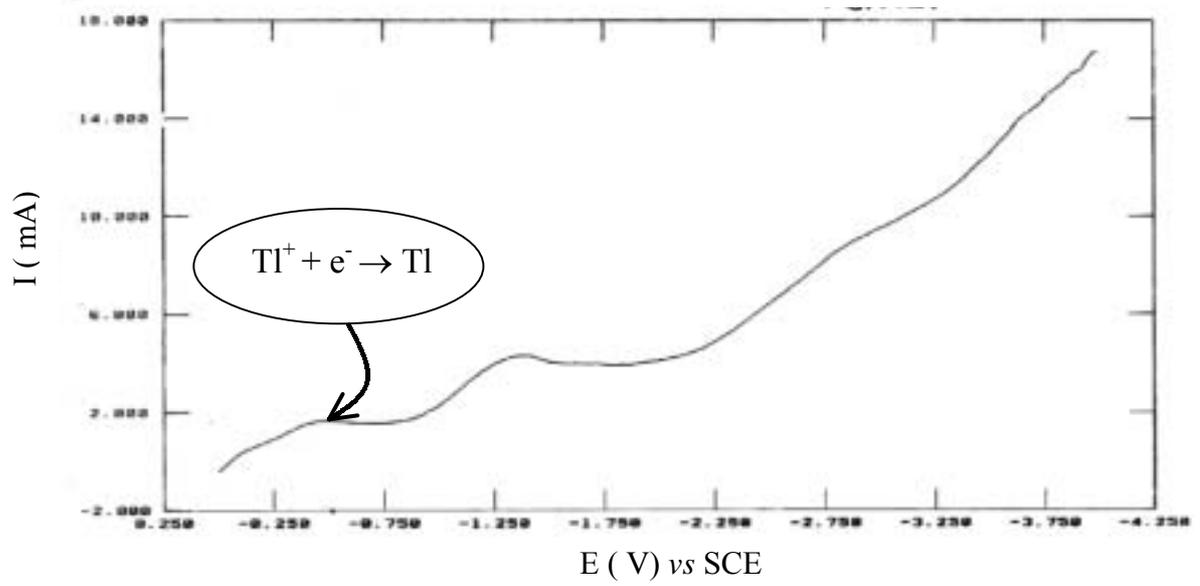

**Figure 6**    Linear sweep voltammogram for the deposition of
Tl-Ba-Ca-Cu alloy onto Ag/CdSe film.

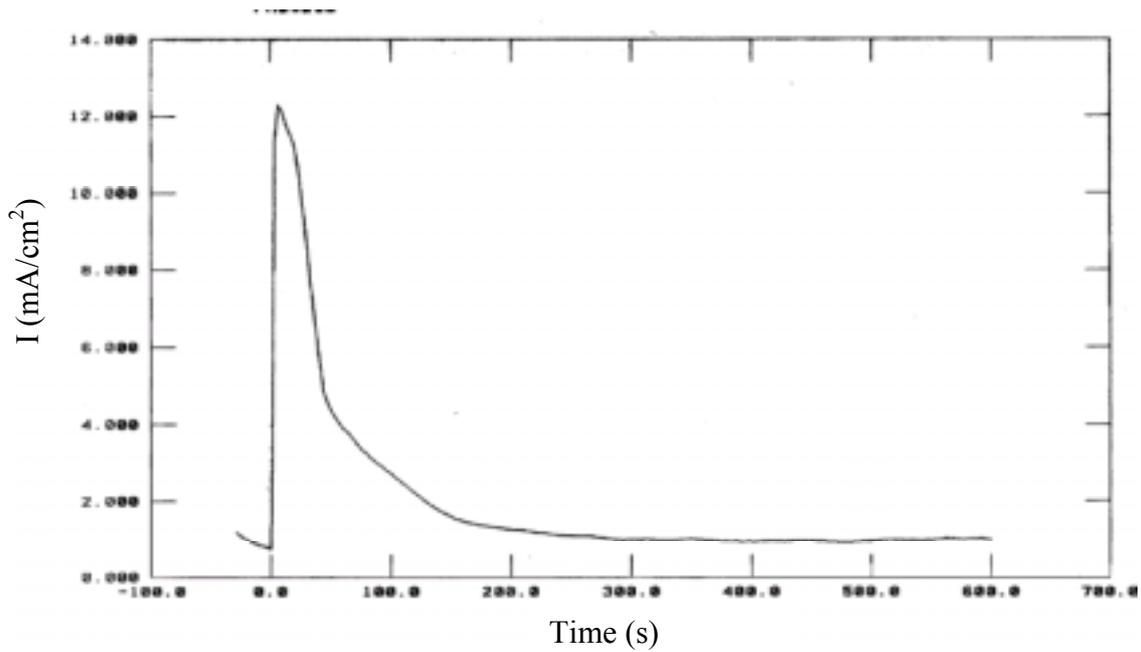

**Figure 7** (a)  Chronoamperometry  Plot for the deposition of the
Tl-Ba-Ca-Cu alloy  deposition onto Ag/CdSe film.

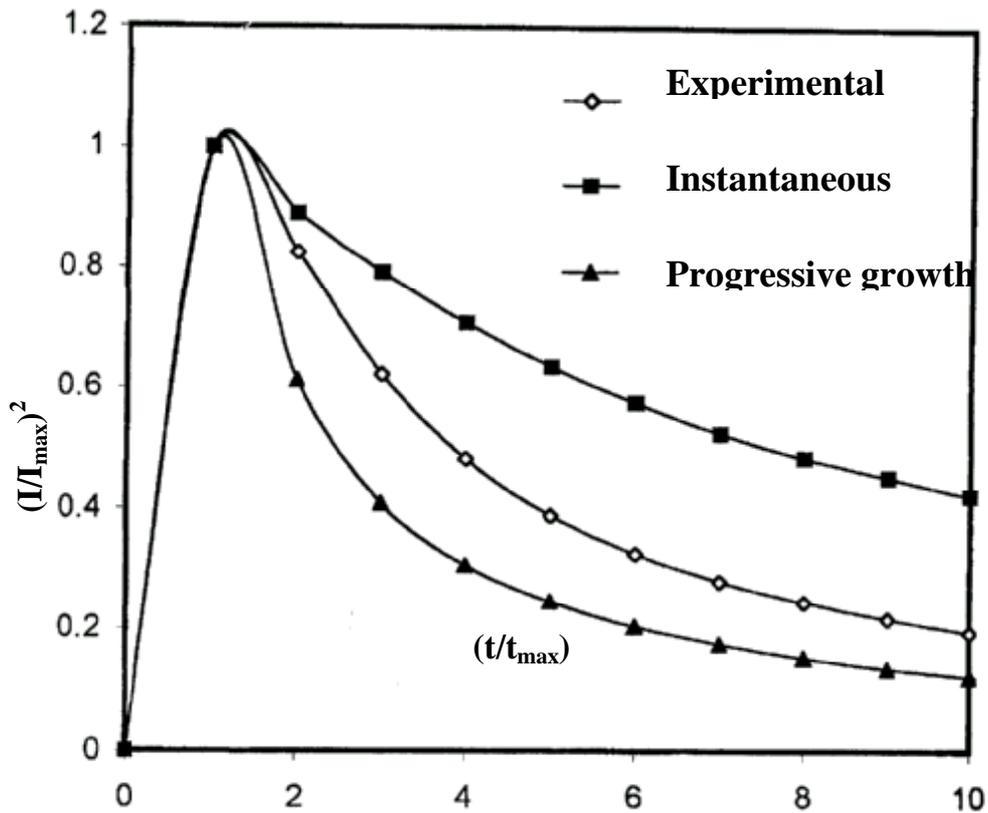

**Figure 7** (b)    The plot of  $(I/I_{max})^2$ *vs* $(t/t_{max})$ for the study of nucleation and
growth mechanism during the deposition of Tl-Ba-Ca-Cu
alloy onto Ag/CdSe film.

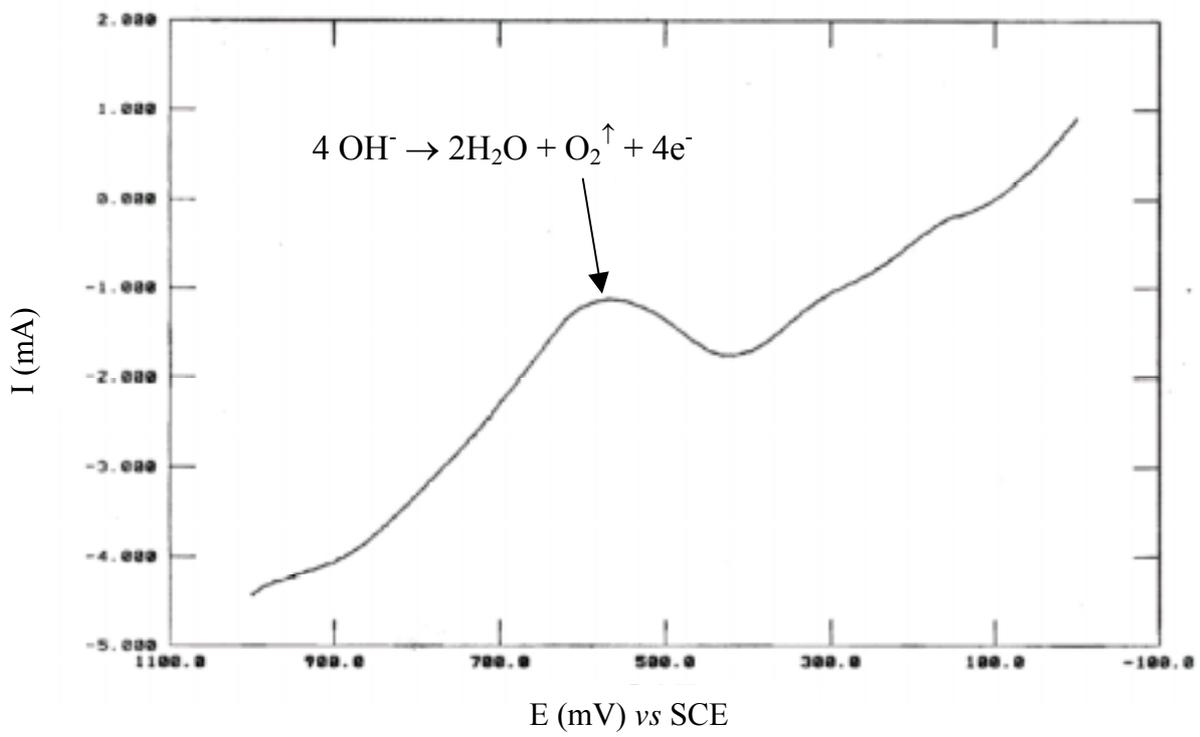

**Figure 8** Linear sweep voltammogram during the intercalation of oxygen into Tl-Ba-Ca-Cu alloy deposited onto CdSe film.

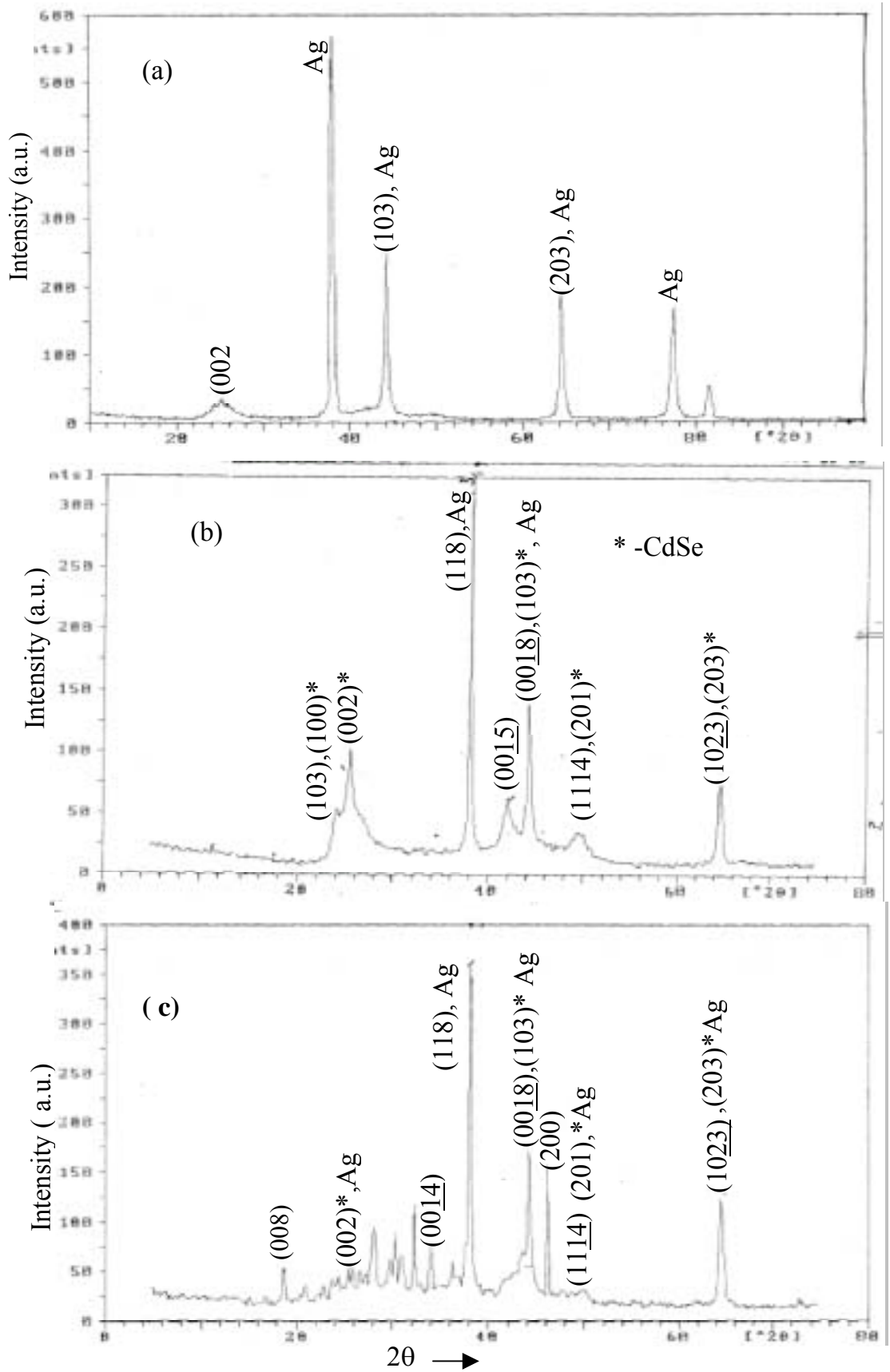

**Figure 9** XRD pattern of the (a) CdSe deposited onto Ag-substrate
(b) Ag/Tl-2223/CdSe heterostructure (c) Ag/CdSe/Tl-2223 heterostructure

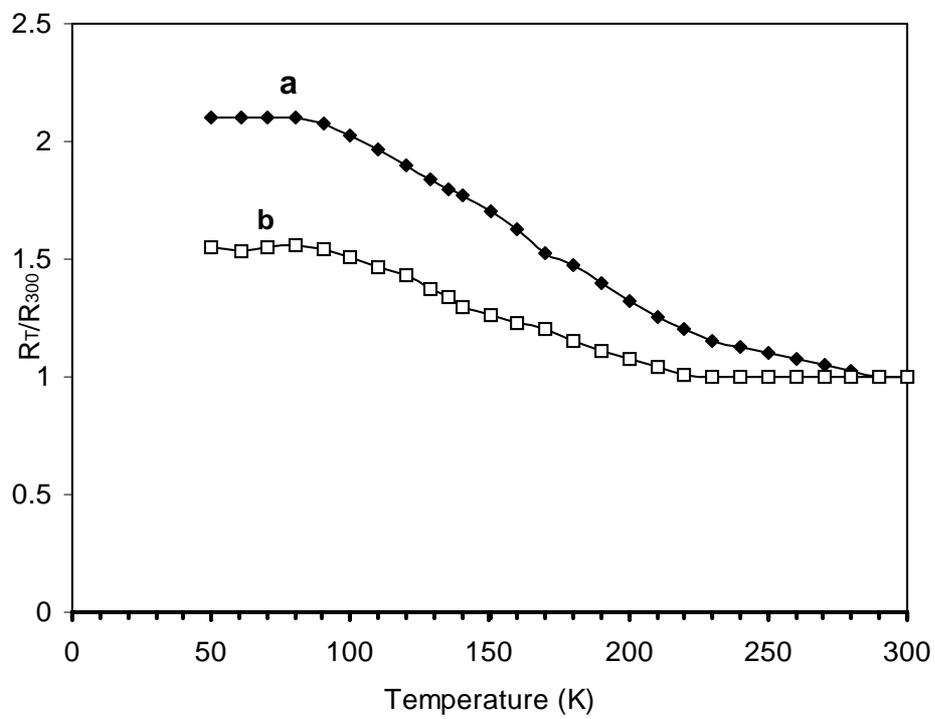

**Figure 11**   Variation of the normalized resistance with
the temperature of Ag/Tl-2223/CdSe
heterostructure (a) Before laser irradiation
and (b) After laser irradiation.

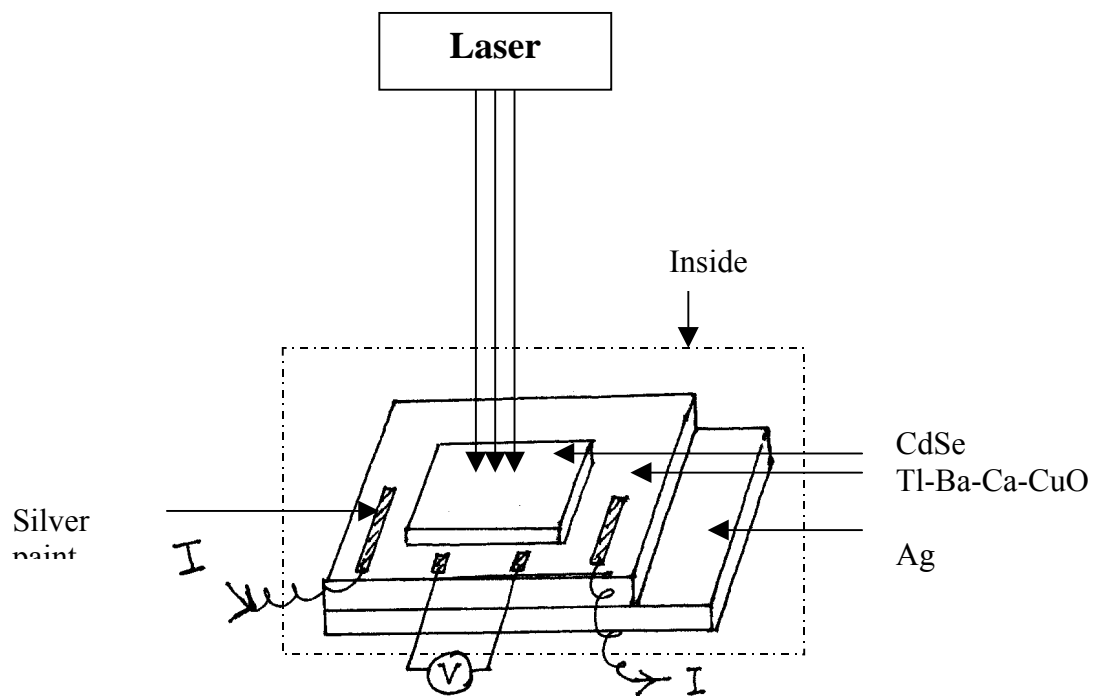

**Figure 12**   The geometry of Ag/Tl-2223/CdSe heterostructure and contacts made for the resistivity measurement of Tl-Ba-Ca-CuO films

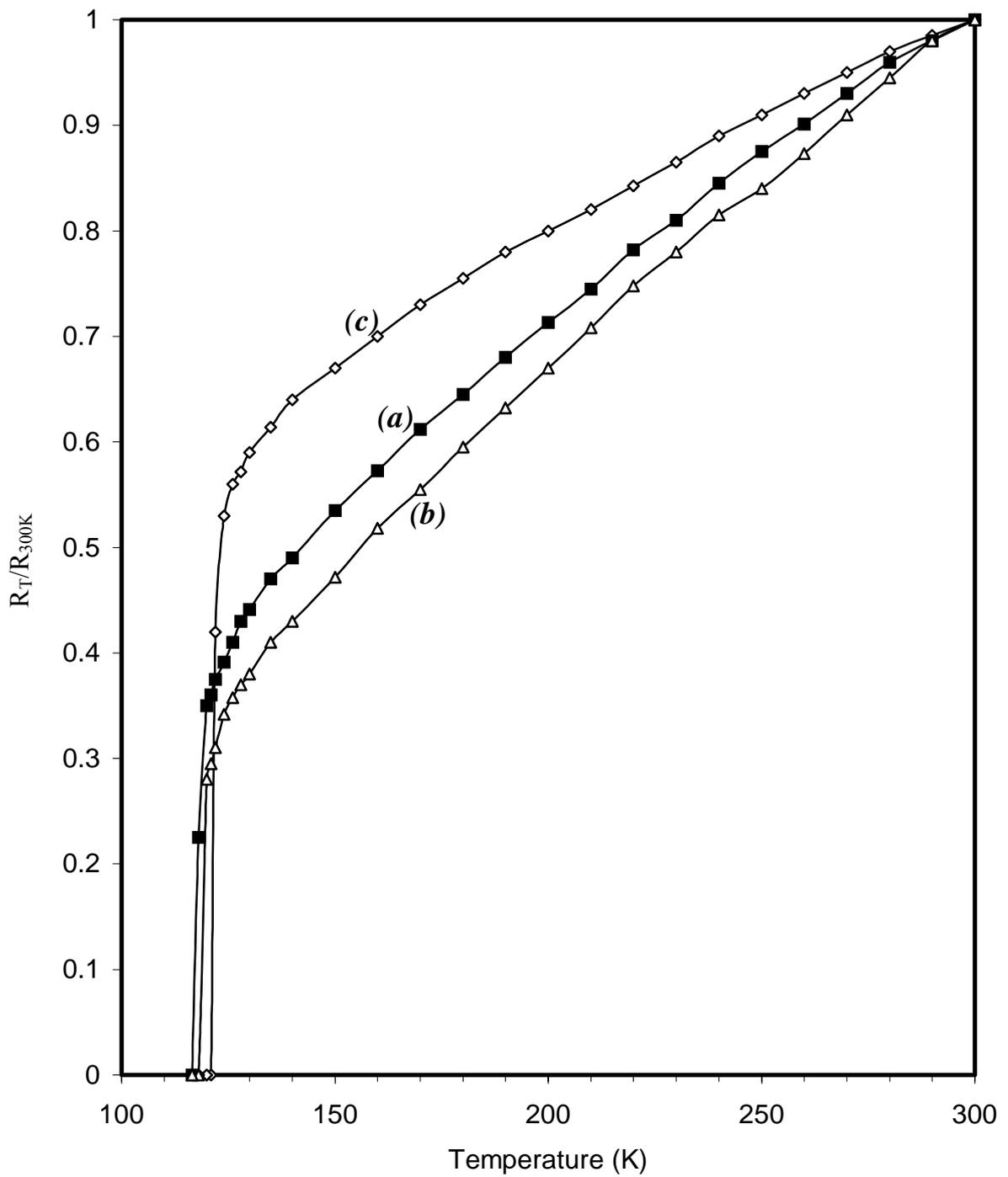

**Figure 13**     Variation in the normalized resistance with the temperature of
(a) Tl-2223 film
(b) Tl-2223 film after CdSe deposition and
(c) Tl-2223/CdSe after laser irradiation

(a)

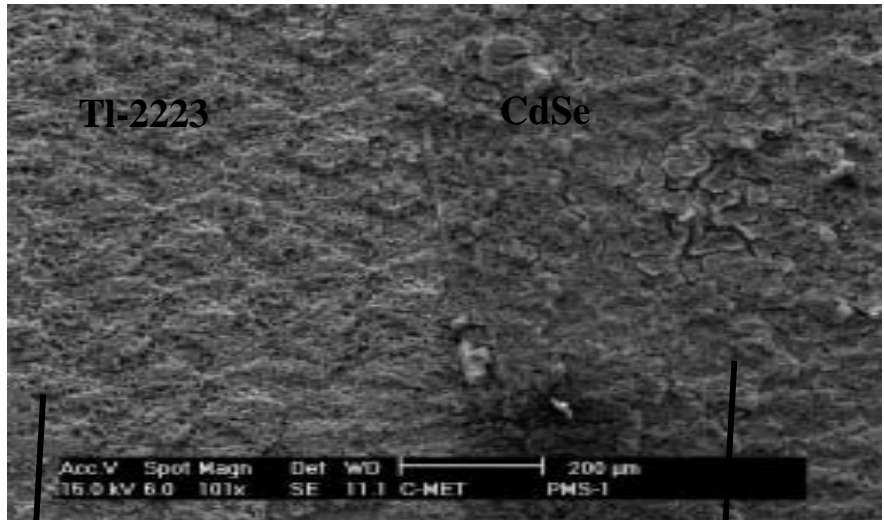

Tl-2223      CdSe

(b)

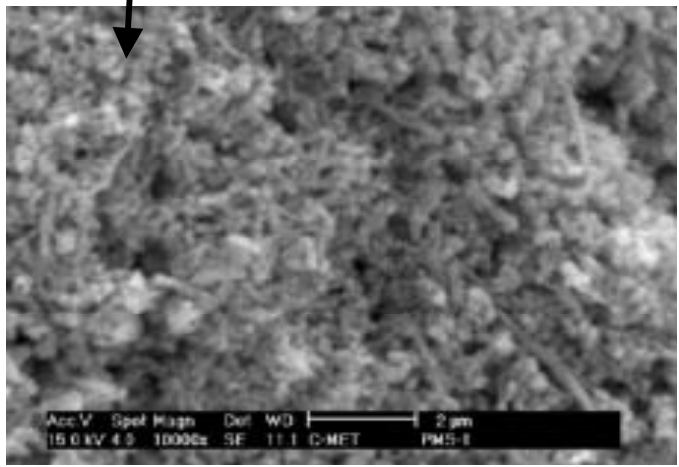

(c)

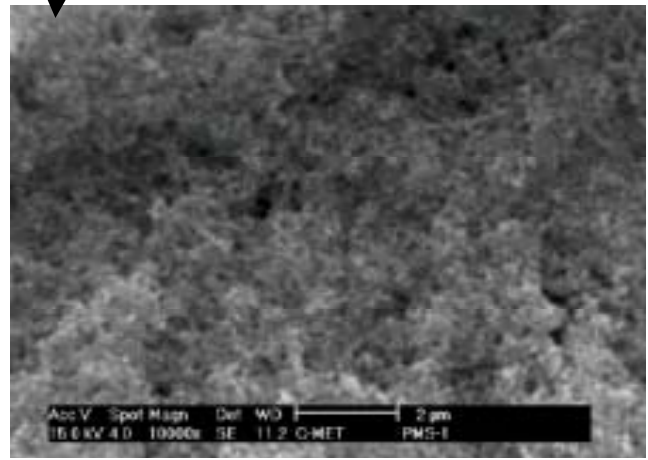

**Figure 10**